\documentclass[preprint,superscriptaddress,nofootinbib,onecolumn,aps,prd]{revtex4-1}
\usepackage{amsmath}
\usepackage{slashbox}
\usepackage{graphicx}
\usepackage{caption}
\usepackage{subcaption}
\usepackage{fixme}
\usepackage{tikz}
\usetikzlibrary{trees,arrows}
\usetikzlibrary{decorations.pathmorphing}
\usetikzlibrary{decorations.markings}

\newcommand{\p}{\partial}
\newcommand{\Int}{\int\limits}

\newcommand{\e}{\mathrm{e}}
\newcommand{\Br}{\mathrm{Br}}
\newcommand{\lagrangian}{\mathcal{L}}

\tikzset{
 >=stealth',
every node/.style={font=\scriptsize},
   photon/.style={decorate, decoration={snake,amplitude=2pt,segment
       length=8pt,pre length=0cm,post length=0}},
   photontiny/.style={decorate, decoration={snake,amplitude=1pt,segment
       length=4pt,pre length=0cm,post length=0}},
   electron/.style={postaction={decorate},
       decoration={markings,mark=at position .55 with {\arrow{>}}}},
   antielectron/.style={postaction={decorate},
       decoration={markings,mark=at position .52 with {\arrow{<}}}},
   drelectron/.style={line width=1.5pt,postaction={decorate},
       decoration={markings,mark=at position .52 with {\arrow{>}}}},
   gluon/.style={decorate,
       decoration={coil,amplitude=4pt, segment length=5pt}},
   scalar/.style={densely dashed}
} 

\fxsetup{layout=inline}
\renewcommand{\fxnotename}[1]{}

\begin{document}

\title{Extending the Higgs sector: an extra singlet.}

\author{S.I. Godunov}
\email{sgodunov@itep.ru}
\affiliation{Institute for Theoretical and Experimental Physics, Moscow, 117218, Russia}
\affiliation{Novosibirsk State University, Novosibirsk, 630090, Russia}
\author{A.N. Rozanov}
\affiliation{The Center for Particle Physics of Marseilles, Marseille, F-13288,
France}
\author{M.I. Vysotsky}
\email{vysotsky@itep.ru}
\affiliation{Institute for Theoretical and Experimental Physics, Moscow, 117218, Russia}
\affiliation{Moscow Institute of Physics and Technology, 141700, Dolgoprudny, Moscow
Region, Russia}
\affiliation{Moscow Engineering Physics Institute, 115409, Moscow, Russia}
\author{E.V. Zhemchugov}
\email{zhemchugov@itep.ru}
\affiliation{Institute for Theoretical and Experimental Physics, Moscow, 117218, Russia} 
\affiliation{Moscow Engineering Physics Institute, 115409, Moscow, Russia}

\begin{abstract}
 An extension of the Standard Model with an additional Higgs singlet is
 analyzed. Bounds on singlet admixture in 125~GeV $h$ boson from electroweak
 radiative corrections and data on $h$ production and decays are obtained.
 Possibility of double $h$ production enhancement at 14~TeV LHC due to heavy
 higgs contribution is considered.
\end{abstract}

\maketitle

\listoffixmes

\section{Introduction}

After the discovery of the Higgs (BEH) boson~\cite{higgs-atlas,higgs-cms}, all
fundamental particles of the Standard Model (SM) are finally found, and now even
passionate adepts of the SM should look for physics beyond it. The pattern of
particles we have is rather asymmetric: there are twelve vector bosons, many
leptons and quarks with spin $1/2$ and only one scalar particle $h$ with mass
125~GeV. Of course, there is only one particle with spin 2 as well, a graviton.
However, unlike the spin 2 case, there are no fundamental principle according to
which there should exist only one fundamental scalar particle. That is why it is
quite probable that there are other still undiscovered fundamental scalar
particles in Nature. The purpose of the present paper is to consider the
simplest extension of the SM by adding one real scalar field to it.  Such an
extension of the SM attracts considerable attention: relevant references can be
found in recent papers~\cite{dawson,robens,martin-lozano,lebedev}. Extra
singlet can provide first order electroweak phase transition needed for
electroweak baryogenesis. It can act as a particle which connects SM particles
to Dark Matter. Not going into these (very interesting) applications, we will
study the degree of enhancement of double higgs production at LHC due to an
extra singlet. To do this we should analyze bounds on the mass of the additional
scalar particle and its mixing with isodoublet state.

An enhancement of $hh$ production occurs due to the mixing of the SM isodoublet
with additional scalar field which is proportional to the vacuum expectation
value (vev) of this field. Thus isosinglet is singled out: its vev does not
violate custodial symmetry and can be large. For higher representations special
care is needed; see paper~\cite{isotriplet} where an introduction of
isotriplet(s) in the SM is discussed.

The paper is organized as follows: in Section~\ref{s:model} we describe the
model and find the physical states. In Section~\ref{s:sm-lhc} we get bounds on
the model parameters of the scalar sector from the experimental data on $h$
production and decays and from precision measurements of $Z$- and $W$-boson
parameters and $t$-quark and $h$ masses. In Section~\ref{s:production} we
discuss double $h$ production at LHC Run~2. In
Appendix~\ref{s:effective-lagrangian} qualititative description of single and
double higgs production at LHC is presented.

\section{The model}

\label{s:model}

Adding to the SM a real field $X$, we take the scalar fields potential in the
following form:
\begin{equation}
 V(\Phi, X)
 = -\frac{m_\Phi^2}{2} \Phi^\dagger \Phi
 + \frac{m_X^2}{2} X^2
 + \frac{\lambda}{2} (\Phi^\dagger \Phi)^2
 + \mu \Phi^\dagger \Phi X,
 \label{potential}
\end{equation}
where $\Phi$ is an isodoublet.\footnote{ We are grateful to J.~M.~Fr\`ere who
 brought to our attention that similar model was considered long ago
in~\cite{hill}.}  Terms proportional to $X^3$, $X^4$ and $\Phi^\dagger \Phi X^2$
are omitted despite that they are allowed by the demand of renormalizability: we
always may assume that they are multiplied by small coupling constants. Two
combinations of the parameters entering~\eqref{potential} are known
experimentally: it is the mass of one of the two scalar states, $h$, which
equals 125~GeV and the isodoublet expectation value $v_\Phi = 246$~GeV. The two
remaining combinations are determined by the mass of the second scalar, $H$ (we
take $m_H > m_h$, though this is not obligatory), and the angle $\alpha$ which
describes singlet-doublet admixture:
\begin{equation}
 \left\{
  \begin{aligned}
   h &= \phi \cos \alpha + \chi \sin \alpha, \\
   H &= -\phi \sin \alpha + \chi \cos \alpha,
  \end{aligned}
 \right.
 \qquad
 \left\{
  \begin{aligned}
   \phi &= h \cos \alpha - H \sin \alpha, \\
   \chi &= h \sin \alpha + H \cos \alpha.
  \end{aligned}
 \right.
\end{equation}
Substituting in~\eqref{potential}
\begin{equation}
 \Phi = \begin{pmatrix}
  \phi^+ \\
  \frac{1}{\sqrt{2}} (v_\Phi + \phi + i \eta)
 \end{pmatrix},
 \qquad
 X = v_X + \chi,
\end{equation}
at the minimum of the potential we get:
\begin{equation}
 \left\{
  \begin{aligned}
   \lambda v_\Phi^2 + 2 \mu v_X &= m_\Phi^2, \\
     2 m_X^2 v_X + \mu v_\Phi^2 &= 0,
  \end{aligned}
 \label{potential-minimum}
 \right.
\end{equation}
so $\mu$ is negative. For the mass matrix using~\eqref{potential-minimum} we
get:
\begin{equation}
 M = \begin{pmatrix}
      V_{\phi \phi} & V_{\phi \chi} \\
      V_{\phi \chi} & V_{\chi \chi}
     \end{pmatrix}
 = \begin{pmatrix}
    \lambda v_\Phi^2 & \mu v_\Phi \\
    \mu v_\Phi       & m_X^2
   \end{pmatrix},
 \label{mass-matrix}
\end{equation}
where $V_{\phi \chi} \equiv \frac{\p^2 V}{\p \phi \p \chi}, \ldots$ Eigenvalues
of~\eqref{mass-matrix} determine masses of scalar particles:
\begin{equation}
 m_{h,H}^2
 =   \frac{1}{2} \lambda v_\Phi^2 + \frac{1}{2} m_X^2
 \mp \sqrt{
      \left(\frac{1}{2} \lambda v_\Phi^2 - \frac{1}{2} m_X^2 \right)^2 + \mu^2 v_\Phi^2
     },
 \label{masses}
\end{equation}
where ``$-$'' corresponds to $m_h$ and ``$+$''---to $m_H$. Eigenfunctions are
determined by the mixing angle~$\alpha$:
\begin{equation}
 \sin 2 \alpha = \frac{-2 \mu v_\Phi}{m_H^2 - m_h^2},
 \quad
 \tan \alpha = \frac{m_h^2 - \lambda v_\Phi^2}{\mu v_\Phi}.
 \label{sine}
\end{equation}

\begin{figure}
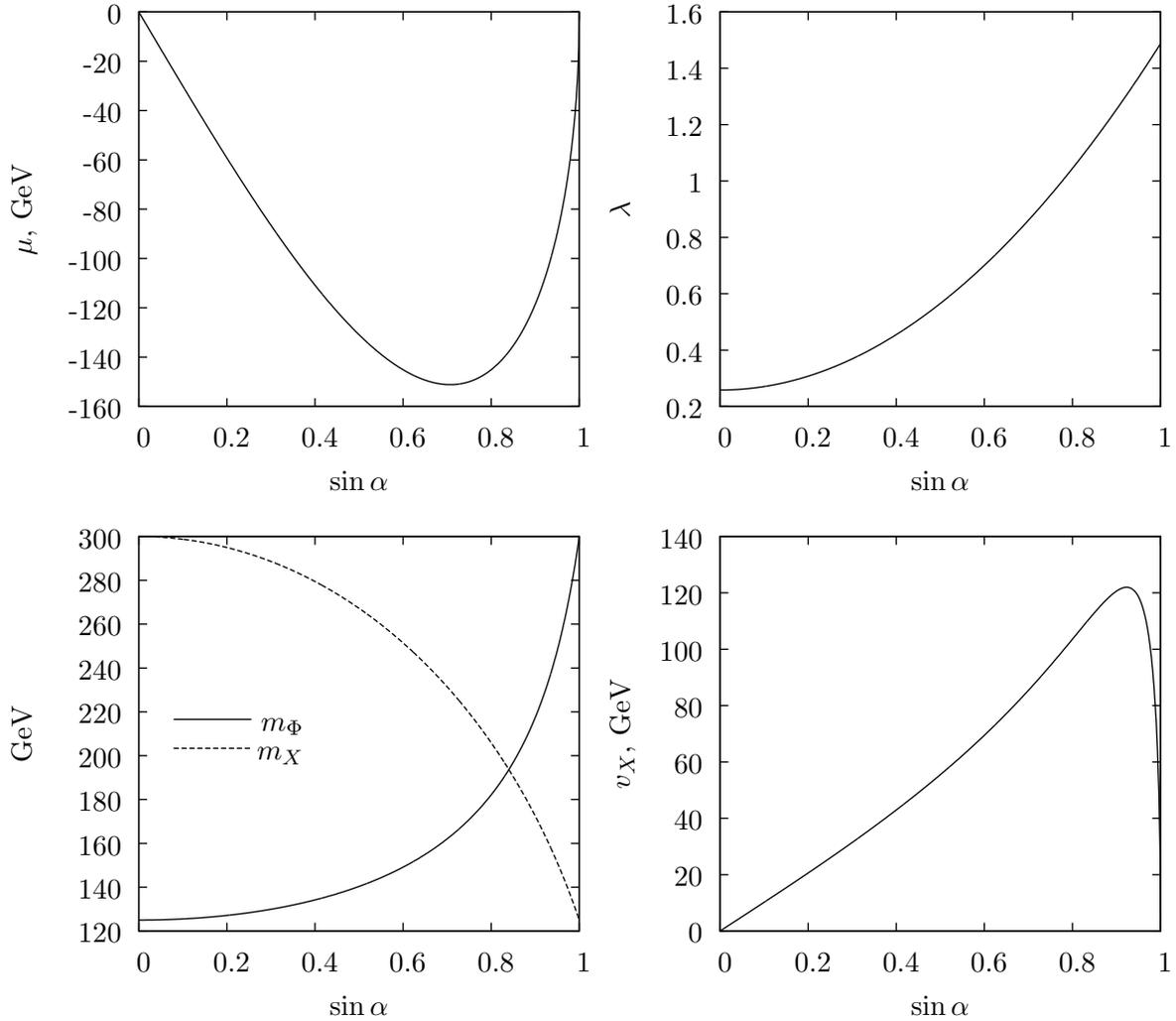

 \centering
 \begin{subfigure}{0.45\textwidth}
  \input{mu.txt}
 \end{subfigure}
 ~
 \begin{subfigure}{0.45\textwidth}
  \input{lambda.txt}
 \end{subfigure}
 \\
 \begin{subfigure}{0.45\textwidth}
  \input{masses.txt}
 \end{subfigure}
 ~
 \begin{subfigure}{0.45\textwidth}
  \input{vX.txt}
 \end{subfigure}
 \caption{Dependencies of the model parameters on the mixing angle for $m_H =
 300$~GeV.}
 \label{fig:parameters}
\end{figure}

Equations~\eqref{sine} determine $\mu$ and $\lambda$ for the given mixing angle
$\alpha$, while equations~\eqref{masses} determine $m_X$ for given $\alpha$ as
well. Finally, equations~\eqref{potential-minimum} determine the values of
$m_\Phi$ and~$v_X$. Fig.~\ref{fig:parameters} demonstrates the dependencies just
described for $m_H = 300$~GeV.

\section{Bounds from $h$ production at LHC and electroweak precision
observables}

\label{s:sm-lhc}

ATLAS and CMS collaborations had detected $h$ production and decays in the
reactions
\begin{equation}
 pp \to h \to f_i,
 \label{reactions}
\end{equation}
where $f_i, i = 1, 2, \ldots, 5$ designate the so-called ``Big five'' final
state channels: $WW^*$, $ZZ^*$, $\gamma \gamma$, $\tau \bar \tau$, $b \bar b$.
Cross sections of reactions~\eqref{reactions} are equal to the higgs production
cross section times branching ratio of the corresponding decay channel.
Quantities $\mu_i$ are introduced according to the following definition:
\begin{equation}
 \mu_i
 \equiv \frac{
  \sigma_{pp \to h} \cdot \Gamma_{h \to f_i} / \Gamma_h
 }{
  (\sigma_{pp \to h} \cdot \Gamma_{h \to f_i} / \Gamma_h)_\text{SM}
 }.
\end{equation}
According to ATLAS and CMS results, all $\mu_i$ are compatible with one within
experimental and theoretical accuracy. It means that no New Physics are up to
now observed in $h$ production and decays.

In the model with an extra isosinglet, production and decay probabilities of $h$
equal that in the SM multiplied by a factor $\cos^2 \alpha$, that is why we
have:
\begin{equation}
 \mu_i = \cos^2 \alpha,
\end{equation}
and existing bounds on $\mu_i$ are translated into bounds on the mixing angle
$\alpha$. Taking into account all measured production and decay channels, for the
average values experimentalists obtain~\cite{atlas, cms}:
\begin{align}
 \text{ATLAS:} && \mu &= 1.30^{+0.18}_{-0.17}, \label{mu-atlas} \\
 \text{CMS:}   &&
 \mu &=  1.00^{+0.14}_{-0.13} \left[
  \pm 0.09\text{(stat.)}
  ^{+0.08}_{-0.07}\text{(theor.)}
  \pm 0.07\text{(syst.)}
 \right]
 \label{mu-cms}
\end{align}
Let us stress that the theoretical uncertainty in the calculation of $pp \to h$
production cross section at LHC does not allow to reduce substantially the
uncertainty in the value of $\mu$.  Bounds from electroweak precision
observables (EWPO) are not affected by this particular uncertainty. 

We fit experimental data with the help of LEPTOP program~\cite{leptop} using
$m_h = 125.14$~GeV. The result of the SM fit which accounts the $h$ mass
measurement is shown in Table~\ref{tbl:sm-fit}. Quality of the fit is
characterised by the $\chi^2$ value
\begin{equation}
 \chi^2 / n_\text{d.o.f.} = 19.6 / 13.
 \label{sm-chi2}
\end{equation}

\begin{table}
 \centering
 \caption{EWPO fit of the Standard Model}
 \begin{tabular}{|l|c|c|r|}
 \hline
 Observable      & Experimental value & Standard Model & Pull \hspace{1ex}     \\ \hline
 $\Gamma_Z$, GeV & $2.4952(23)$ & $2.4966(14)$ & $-0.5895$ \\ \hline
 $\sigma_h$, nb & $41.541(37)$ & $41.475(14)$ & $1.7746$ \\ \hline
 $R_l$ & $20.771(25)$ & $20.744(18)$ & $1.0831$ \\ \hline
 $A_\text{FB}^l$ & $0.0171(10)$ & $0.0165(2)$ & $0.6572$ \\ \hline
 $A_\tau$ & $0.1439(43)$ & $0.1484(7)$ & $-1.0452$ \\ \hline
 $R_b$ & $0.2163(7)$ & $0.2158(0)$ & $0.7699$ \\ \hline
 $R_c$ & $0.1721(30)$ & $0.1722(0)$ & $-0.0277$ \\ \hline
 $A_\text{FB}^b$ & $0.0992(16)$ & $0.1040(5)$ & $-3.0303$ \\ \hline
 $A_\text{FB}^c$ & $0.0707(35)$ & $0.0744(4)$ & $-1.0565$ \\ \hline
 $s_l^2$ $(Q_\text{FB})$ & $0.2324(12)$ & $0.2313(1)$ & $0.8771$ \\ \hline
 $A_\text{LR}$ & $0.1514(22)$ & $0.1484(7)$ & $1.3822$ \\ \hline
 $A_b$ & $0.923(20)$ & $0.9349(1)$ & $-0.5941$ \\ \hline
 $A_c$ & $0.670(27)$ & $0.6685(3)$ & $0.0567$ \\ \hline
 $M_W$, GeV & $80.3846(146)$ & $80.3725(67)$ & $0.8322$ \\ \hline
 $m_t$, GeV & $173.24(95)$ & $174.32(89)$ & $-1.1370$ \\ \hline
 $1 / \bar \alpha$ & $128.954(48)$ & $129.023(37)$ & $-1.4378$ \\ \hline
\end{tabular}

 \label{tbl:sm-fit}
\end{table}

\begin{figure}[p]
 \centering
 \begin{subfigure}{\textwidth}
  \def\svgwidth{0.56\textwidth}
  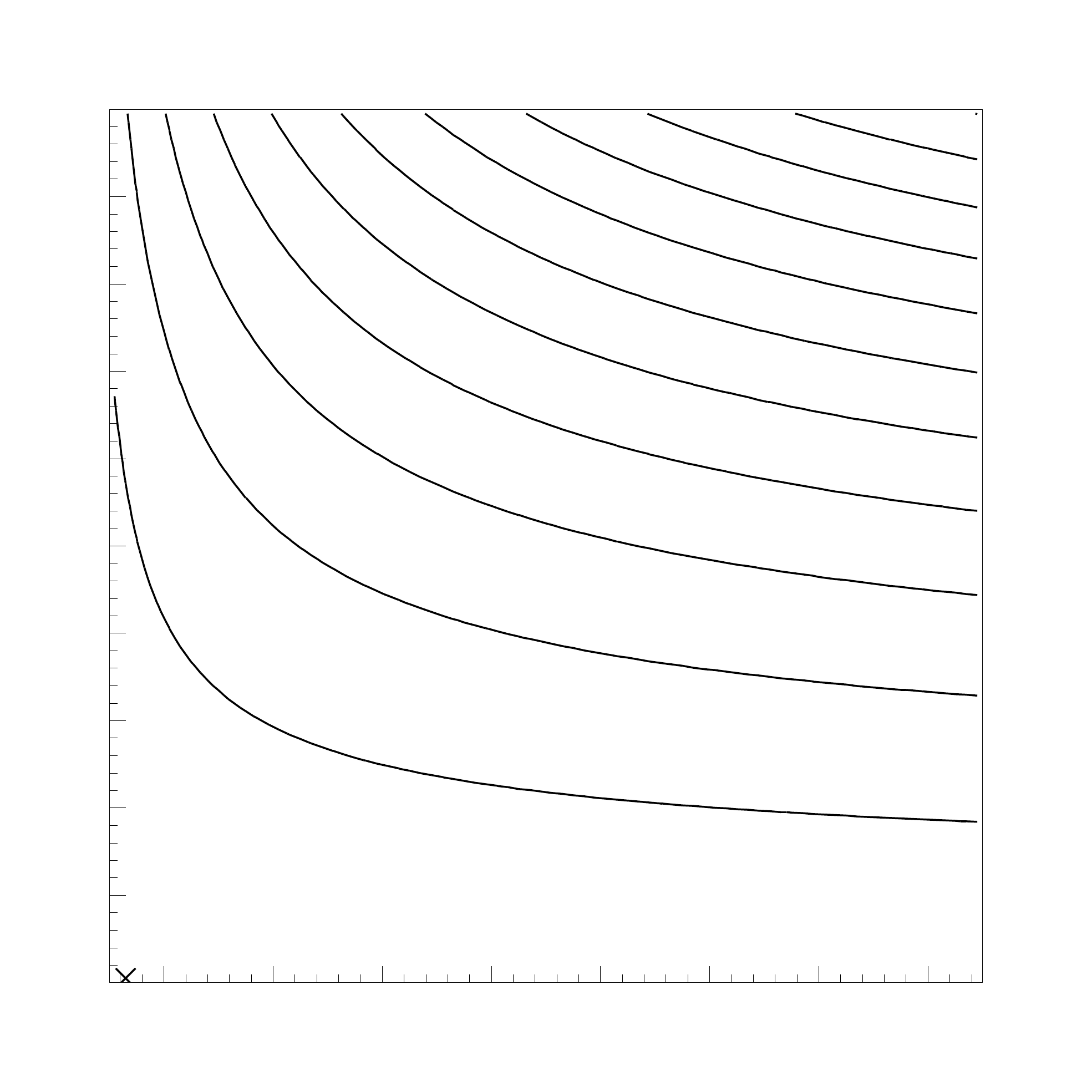
  \caption{Bounds from electroweak precision observables.}
  \label{fig:ew-bounds}
 \end{subfigure}
 \\
 \begin{subfigure}{\textwidth}
  \def\svgwidth{0.56\textwidth}
  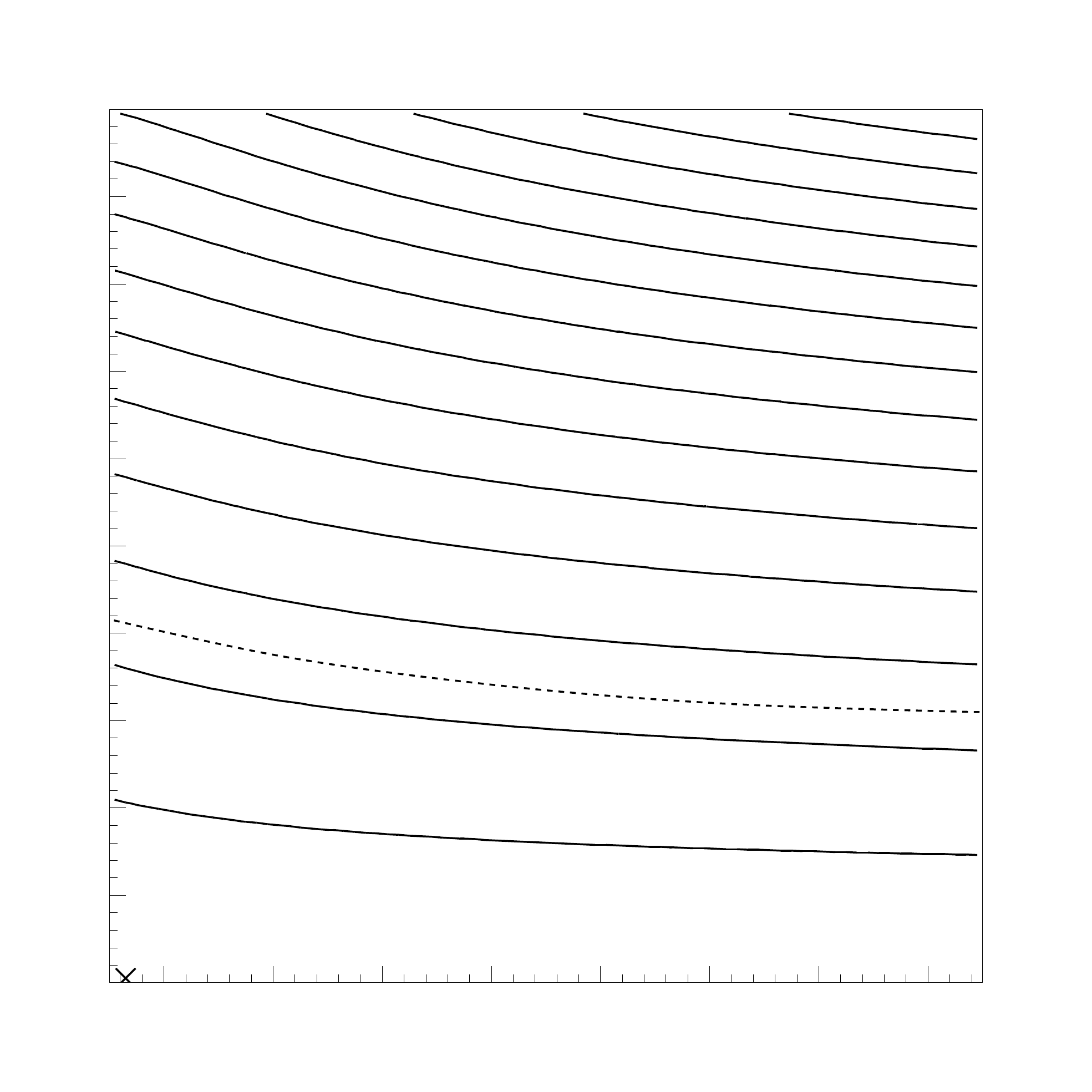
  \caption{Bounds from both electroweak precision observables and signal
   strength measurements~\eqref{mu-atlas},~\eqref{mu-cms}. The dashed line
   corresponds to $\Delta \chi^2 = 5.99$; the probability that numerical values
   of $(m_H, \sin \alpha)$ are below it equals 95\% (compare with
   Ref.~\cite{lebedev}, eq. (23)).
  }
  \label{fig:ew+mu-bounds}
 \end{subfigure}
 \caption{Bounds on the singlet model parameters.}
 \label{fig:ew-and-ew+mu-bounds}
\end{figure}

Higgs boson contributions to electroweak observables at one loop are described
in LEPTOP by functions $H_i(h) = H_i(m_h^2 / m_Z^2)$. In the case of an extra
singlet the following substitution should be performed:
\begin{equation}
 H_i(h) \to \cos^2 \alpha \; H_i(h) + \sin^2 \alpha \; H_i(H),
 \ H = m_H^2 / m_Z^2.
\end{equation}
The same substitution should be made for the functions $\delta_4 V_i(t, h)$, $t
= m_t^2 / m_Z^2$, which describe two loops radiative corrections enhanced as
$m_t^4$. In two loops quadratic dependence on higgs mass appears which is
described by functions $\delta_5 V_i$. Calculations of these corrections in the
case of an extra singlet higgs is not easy. An approximate upper bound has been
estimated by assuming that
\begin{equation}
 \delta_5 V_i(H)
 < \delta_5 V_i((1000 \text{ GeV})^2 / m_Z^2)
 \approx 100 \; \delta_5 V_i(h)
 \text{ for } m_H < 1000 \text{ GeV.}
\end{equation}
Comparison of two calculations, one with $\delta_5 V_i(h) = \cos^2 \alpha \;
\delta_5 V_i(h)$, and the other with
\begin{equation}
 \delta_5 V_i(h)
 = \cos^2 \alpha \; \delta_5 V_i(h)
 + 100 \cdot \sin^2 \alpha \; \delta_5 V_i(h),
\end{equation}
showed that the correction to the values of $\sin \alpha$ in
Fig.~\ref{fig:ew-and-ew+mu-bounds} is less than $10^{-3}$.

Bounds from EWPO on the singlet model parameters are presented in
Fig.~\ref{fig:ew-bounds}. $\chi^2$ minimum is reached at $\sin \alpha = 0$, $m_H
= 150$~GeV, which is the minimum value allowed for $m_H$ in the fit.
Experimental data are avoiding heavy higgs. The value of $\chi^2$ at the minimum
coincides with the SM result~\eqref{sm-chi2}. Lines of constant $\chi^2$
correpospond to $\Delta \chi^2 = 1, 4, 9, \ldots$. Probabilities that $(\sin
\alpha, m_H)$ values are below these lines are 39\%, 86\%, $98.9\%$,
\ldots.\footnote{
 Let us note that if a subset of experimental data from Table~\ref{tbl:sm-fit}
 is fitted, then allowed domains of the $(\sin \alpha, m_H)$ values will be
 larger than those presented in Fig.~\ref{fig:ew-bounds}.  Here we disagree with
 the statement made in~\cite{robens} that the fit of only one observable ($m_W$)
 allows to set the strongest constraint on $(\sin \alpha, m_H)$.
}

Bounds accounting for both EWPO and direct $h$ production
data~\eqref{mu-atlas},~\eqref{mu-cms} are shown in Fig.~\ref{fig:ew+mu-bounds}.
We see that for heavy $H$ bounds from EWPO dominate, while for light $H$
measurement of $\mu$ is more important.

\section{$h$, $H$ and $hh$ production at LHC}

\label{s:production}

The main purpose of this section is to find what enhancement of double higgs
production cross section is possible with enlarged higgs sector. Let us remind
that in the SM double $h$ production cross section is very small. According to
the recent result~\cite{florian}, at $\sqrt{s} = 14$~TeV $\sigma^\text{NNLO}(pp
\to hh) = 40$~fb with a $10 \div 15\%$ accuracy. We will demonstrate that
enlarged higgs sector allows to strongly enhance double $h$ production.

The cross section of $H$ production at LHC equals that for the SM higgs
production (for $(m_h)_\text{SM} = m_H$) multiplied by $\sin^2 \alpha$. Cross
section of the SM higgs production at NNLO we take from Table~3
of~\cite{handbook}. In order to obtain cross section of resonant $hh$ production
in $H$ decays we should multiply cross section of $H$ production by $\Br(H \to
hh)$.

\begin{figure}[t]
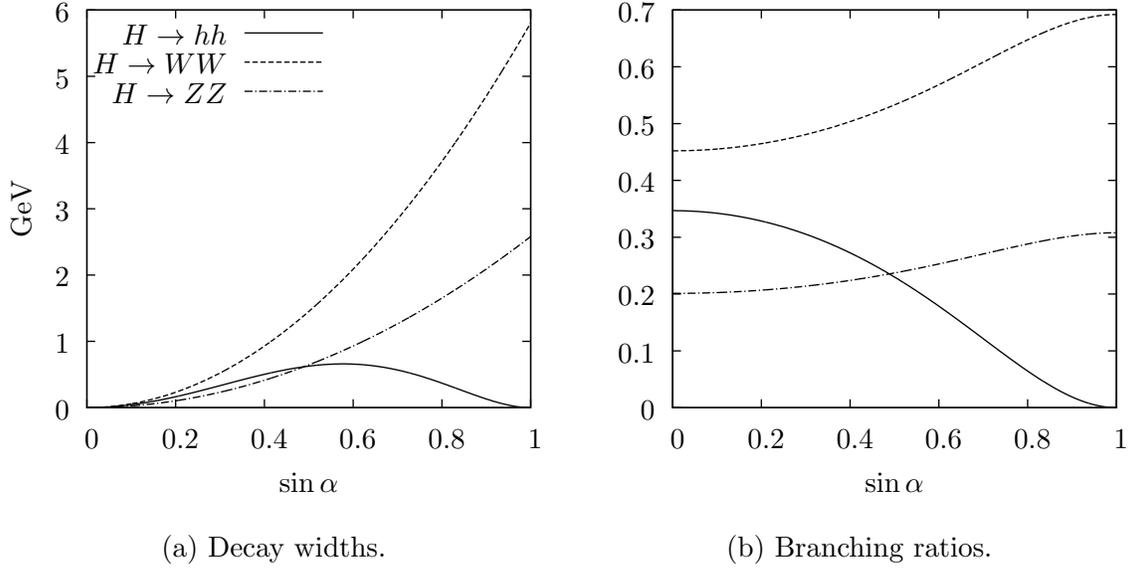

 \centering
 \begin{subfigure}{0.45\textwidth}
  \input{widths.txt}
  \caption{Decay widths.}
 \end{subfigure}
 ~
 \begin{subfigure}{0.45\textwidth}
  \input{branches.txt}
  \caption{Branching ratios.}
  \label{fig:branches}
 \end{subfigure}
 \caption{Decay widths and branching ratios of the heavy higgs boson for $m_H =
  300$~GeV.
 }
 \label{fig:decays}
\end{figure}

Let us consider $H$ decays. Decays to $hh$, $W^+ W^-$, $ZZ$ and $t \bar t$
dominate. For the $Hhh$ coupling we obtain:
\begin{equation}
 \begin{split}
  \Delta \lagrangian_{Hhh}
  &= \left[
       \frac{3}{2} \lambda v_\Phi \cos^2 \alpha \sin \alpha
     - \frac{\mu}{2} \cos \alpha (1 - 3 \sin^2 \alpha)
    \right] H h^2
  \\
  &=    \frac{2 m_h^2 + m_H^2}{2 v_\Phi} \sin \alpha \cos^2 \alpha \; H h^2
  \\
  &\equiv g_{Hhh} H h^2,
 \end{split}
\end{equation}
thus
\begin{equation}
 \Gamma_{H \to hh}
 = \frac{g_{Hhh}^2}{8 \pi m_H} \sqrt{1 - \left( \frac{2 m_h}{m_H} \right)^2}.
\end{equation}
Decays to $W^+ W^-$, $ZZ$, $t \bar t$ occur through isodoublet admixture in $H$:
\begin{equation}
 \begin{split}
  \Delta \lagrangian
  &=      \frac{2 m_W^2}{v_\Phi} \sin \alpha \  H W^+ W^-
   +      \frac{m_Z^2}{v_\Phi}   \sin \alpha \  H Z^2
   +      \frac{m_t}{v_\Phi}     \sin \alpha \  H t \bar t
  \\
  &\equiv g_{HWW} H W^+ W^-
   +      \frac{1}{2} g_{HZZ} H Z^2
   +      g_{H t \bar t} H t \bar t,
 \end{split}
\end{equation}
thus
\begin{align}
 \Gamma_{H \to W^+ W^-}
 &= \frac{g_{HWW}^2 m_H^3}{64 \pi m_W^4}
    \left[ 1 - 4 \frac{m_W^2}{m_H^2} + 12 \frac{m_W^4}{m_H^4} \right]
    \sqrt{1 - \left( \frac{2 m_W}{m_H} \right)^2},
 \\
 \Gamma_{H \to ZZ}
 &= \frac{g_{HZZ}^2 m_H^3}{128 \pi m_Z^4}
    \left[ 1 - 4 \frac{m_Z^2}{m_H^2} + 12 \frac{m_Z^4}{m_H^4} \right]
    \sqrt{1 - \left( \frac{2 m_Z}{m_H} \right)^2},
 \\
 \Gamma_{H \to t \bar t}
 &= \frac{3 g_{Ht\bar t}^2 m_H}{8 \pi}
    \left[ 1 - \left( \frac{2 m_t}{m_H} \right)^2 \right]^\frac{3}{2}.
\end{align}
The dependence of the widths and branching ratios of $H$ decays on mixing angle
$\alpha$ for $m_H = 300$~GeV are shown in Figure~\ref{fig:decays}.

\begin{figure}[p]
 \centering
 \def\svgwidth{0.56\textwidth}
 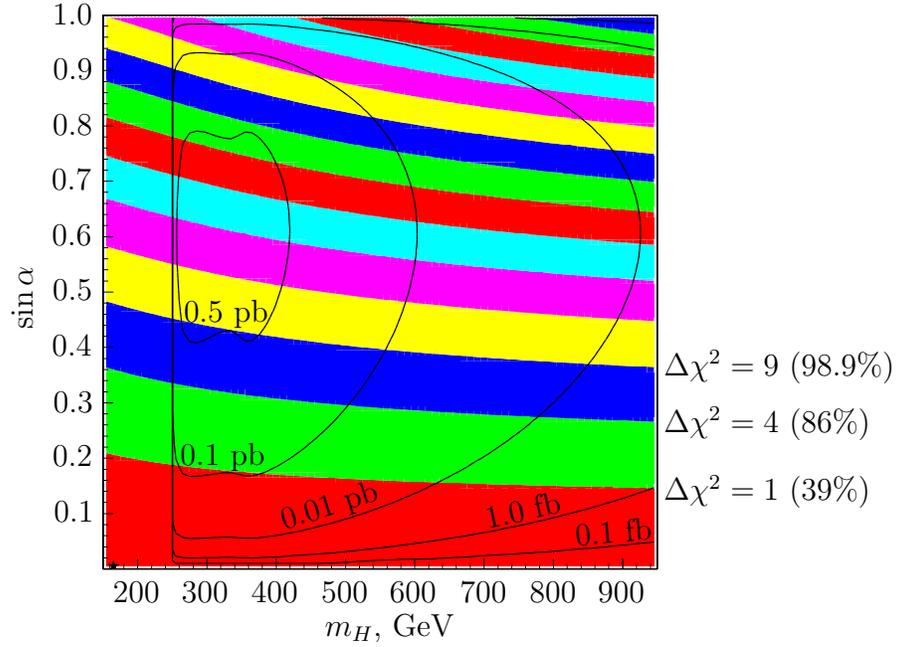
 \caption{
  Contour plot of $\sigma(pp \to H \to hh)$ for $\sqrt{s} = 14$~TeV. \\
  In this figure we neglect small effects of $H \to hh^*$.
 }
 \label{fig:ew-bounds+pp->H->hh-xsection}
\end{figure}

\begin{figure}[p]
 \centering
 \def\svgwidth{0.56\textwidth}
 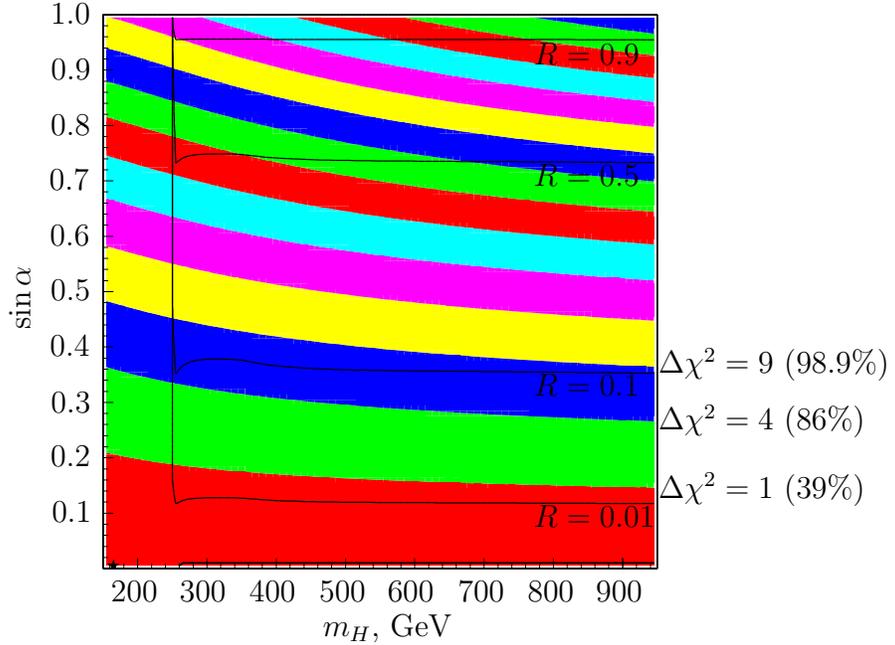
 \caption{
  Contour plot of
  $
   R \equiv \frac{ \sigma(pp \to H) \Br(H \to ZZ)}
                 {(\sigma(pp \to h) \Br(h \to ZZ))_\text{SM}}
  $. \\ In the calculation of $R$ we assume $m_H > 2 m_h$.
 }
 \label{fig:zz-signal}
\end{figure}

For the cross section of the reaction $pp \to H \to hh$ we have:
\begin{equation}
 \sigma(pp \to H \to hh)
 = \sigma(pp \to h)_\text{SM} \cdot \sin^2 \alpha \cdot \Br(H \to hh),
 \label{pp->H->hh-xsection}
\end{equation}
the lines of constant cross section are shown in
Fig.~\ref{fig:ew-bounds+pp->H->hh-xsection} (compare to Fig. 4
from~\cite{lebedev}).  $H \to ZZ$ decay can be used in order to find $H$;
its cross section divided by that for the SM higgs boson with $(m_h)_\text{SM} =
m_H$ is
\begin{equation}
 R
 \equiv \frac{\sigma(pp \to H) \cdot \Br(H \to ZZ)}
             {(\sigma(pp \to h) \cdot \Br(h \to ZZ))_\text{SM}}
 =      \frac{\sin^4 \alpha}
             {\sin^2 \alpha + \frac{\Gamma(H \to hh)}{\Gamma_\text{SM}}}.
\end{equation}
Contour plot of $R$ is presented in Fig.~\ref{fig:zz-signal}. Let us note that
$R$ does not depend on $\sqrt{s}$.

\section{Conclusions}

In the models with extended higgs sector strong resonant enhancement of double
higgs production is possible which makes the search of $pp \to hh$ reaction at
Run~2 LHC especially interesting. According to
Fig.~\ref{fig:ew-bounds+pp->H->hh-xsection} cross section of $pp \to H \to hh$
reaction can be as large as 0.5 pb, ten times larger than the SM value.

The search for $H$ boson can also go in the same way as it was for the heavy SM
boson $h$. Probability of $H$ observation diminishes compared to that of $h$
because of a) suppression of $H$ production cross section by the factor $\sin^2
\alpha \le 0.2$; b) suppression of $\Br(H \to ZZ)$ because of additional $H \to
hh$ decay mode. Taking these two factors into account, we get about factor 10
suppression of $pp \to H \to ZZ$ process probability compared to that for the SM
higgs boson (see Fig.~\ref{fig:zz-signal}).

Results for the search of higgs-like boson in $ZZ$ decay mode can be found
in~\cite{cms-bounds}, Figure~5. Comparing it with our Fig.~\ref{fig:zz-signal},
we observe that experimental data start to be sensitive to the singlet model
expectation for maximally allowed values of the mixing angle $\alpha$.

After the first version of this paper was published in arXiv, we got a number of
emails providing us with references to related research~\cite{followups}.

S.~G., M.~V. and E.~Zh. are partially supported under the grants RFBR
No.~14-02-00995 and NSh-3830.2014.2. S~G. and E.~Zh. are also supported by
MK-4234.2015.2. In addition, S.~G is supported by Dynasty Foundation and by the
Russian Federation Government under grant No.~11.G34.31.0047.

\appendix

\section{Higgs production in effective Lagrangian approach}

\label{s:effective-lagrangian}

Simple analythical formulas which qualititavely describe single and double higgs
production in the SM are presented in this section.  Let us start with single
higgs production in gluon fusion. In the limit $m_h \ll 2 m_t$, the amplitude of
$gg \to h$ transition is determined by the top quark contribution into the QCD
Gell-Mann-Low function:
\begin{equation}
 \Delta \lagrangian 
 = \frac{\alpha_s}{12 \pi} \ln \left( 1 + \frac{h}{v_\Phi} \right) G_{\mu \nu}^2;
 \quad
 M = \frac{\alpha_s}{6 \pi v_\Phi} G_{\mu \nu}^1 G_{\mu \nu}^2 h,
 \label{gg-h-amplitude}
\end{equation}
leading to the well-known result for the production cross section:
\begin{equation}
 \sigma_{gg \to h} 
 = \frac{\alpha_s^2 \tau_0}{576 \pi v_\Phi^2} \delta(\tau - \tau_0).
 \label{gg-h-xsection}
\end{equation}
Here $\tau = \hat s / s$ and $\tau_0 = m_h^2 / s$;
$s \equiv (p_1 + p_2)^2$ is the invariant mass of colliding protons, $\hat s =
x_1 x_2 s \equiv \tau s$ is the invariant mass of colliding gluons. Integrating
over gluons distribution in a proton, we get:
\begin{equation}
 \sigma_{pp \to h}
 = \Int_{\tau_0}^1 dx_1 \Int_{\tau_0 / x_1}^1 dx_2 \ 
     g(x_1) g(x_2) \sigma_{gg \to h}.
 \label{pp-h-xsection}
\end{equation}
Changing the variables from $x_1$, $x_2$ to $\tau$, $y$ according to the
following definitions: $x_1 = \sqrt{\tau} \e^y$, $x_2 = \sqrt{\tau} \e^{-y}$,
and substituting~\eqref{gg-h-xsection} into~\eqref{pp-h-xsection}, we obtain:
\begin{equation}
 \sigma_{pp \to h}
 = \frac{\alpha_s^2 m_h^2}{576 \pi v_\Phi^2} \frac{1}{s}
   \Int_{\ln \sqrt{\tau_0}}^{-\ln \sqrt{\tau_0}}
     g(\sqrt{\tau_0} \e^y) g(\sqrt{\tau_0} \e^{-y}) dy
     \equiv \frac{\alpha_s^2 m_h^2}{576 \pi v_\Phi^2} \frac{dL}{d \hat s},
\end{equation}
where the so-called gluon-gluon luminosity is given by the integral over gluon
distributions:
\begin{equation}
 \left. \frac{dL}{d \hat s} \right\rvert_{\hat s = m_h^2}
 = \frac{1}{s}
   \Int_{\ln \sqrt{\tau_0}}^{-\ln \sqrt{\tau_0}}
     g(\sqrt{\tau_0} \e^y) g(\sqrt{\tau_0} \e^{-y}) dy.
 \label{luminosity}
\end{equation}

A number of PDFs parametrizations exist in the literature; their results
for~\eqref{luminosity} at $\sqrt{s} = 7$, 8, 14 and 100 TeV and $m_h^2 =
(125\text{ GeV})^2$ coincide within several percents. Finite value of $m_t =
172$~GeV should be taken into account by multiplication of the leading order
result for the amplitude $M$~\eqref{gg-h-amplitude} by a factor
\begin{equation}
 F = \frac{3}{2} \beta [(1 - \beta) x^2 + 1],
\end{equation}
where $\beta = \left( \dfrac{2 m_t}{m_h} \right)^2$, and $x = \arctan
\dfrac{1}{\sqrt{\beta - 1}}$ for $\beta > 1$, $x = \dfrac{1}{2} \left( \pi + i \ln
\dfrac{1 + \sqrt{1 - \beta}}{1 - \sqrt{1 - \beta}} \right)$ for $\beta <
1$~\cite{okun} (note that $\lim\limits_{m_t \to \infty} F = 1$).  This
adjustment leads to 6\% enlargement of $\sigma_{gg \to h}$ compared to
$m_t \to \infty$ value; however taking into account $b$ and $c$ quark
contributions results in 6\% overall reduction.

Applying all these factors and using PDFs from~\cite{pdf}, we obtain numbers
presented in Table~\ref{tbl:lhc-data}. To calculate $\sigma^\text{NNLO}$ from
$\sigma^\text{LO}$ we use $K$-factor from~\cite{djouadi}: $K \approx 2.5$ for
$\sqrt{s} = 7$ and 8~TeV, and $K \approx 2$ for $\sqrt{s} = 14$~TeV.
For $\sqrt{s} = 100$~TeV $K \approx 1.5$ (A. Djouadi, private communication).
Let us stress that according to~\cite{djouadi}, accuracy of the calculated value
of $\sigma^\text{NNLO}_{pp \to h}$ is at the level of $\pm (10 \div 17)\%$ which
makes hopes of reducing uncertainty in $\mu_i$ (and $\mu$) below 10\% elusive.
In the case of an extra singlet, $h$ and $H$ production cross sections equal the
SM one multiplied by $\cos^2 \alpha$ and $\sin^2 \alpha$ respectively.

\begin{table}
 \centering
 \caption{Data relevant for the SM higgs boson production at LHC. The difference
  between the numbers in Tables~\ref{tbl:pp->h:NNLO}
  and~\ref{tbl:pp->h:NNLO-handbook} is due to poor accuracy of $K$-factors
  presented in~\cite{djouadi}.
 }
 \begin{subtable}{\textwidth}
  \caption{$\frac{d L}{d \hat s}$, $10^{-3} \text{ GeV}^{-2}$.}
  \begin{tabular}{|c|c|c|c|c|}
   \hline
   \backslashbox{$m_H$}{$\sqrt{s}$} & 7 TeV & 8 TeV & 14 TeV & 100 TeV
   \\ \hline
   125 GeV & 6.41  & 8.30  & 22.9  & 451  \\ \hline
   300 GeV & 0.147 & 0.205 & 0.737 & 25.1 \\ \hline
  \end{tabular}
 \end{subtable}
 \\
 \begin{subtable}{\textwidth}
  \caption{$\sigma^\text{LO}(pp \to h)$, pb.}
  \begin{tabular}{|c|c|c|c|c|}
   \hline
   \backslashbox{$m_H$}{$\sqrt{s}$} & 7 TeV & 8 TeV & 14 TeV & 100 TeV
   \\ \hline
   125 GeV & 5.52  & 7.16 & 19.8 & 389 \\ \hline
   300 GeV & 0.936 & 1.31 & 4.69 & 160 \\ \hline
  \end{tabular}
 \end{subtable}
 \\
 \begin{subtable}{\textwidth}
  \caption{$\sigma^\text{NNLO}(pp \to h)$, pb.}
  \begin{tabular}{|c|c|c|c|c|}
   \hline
   \backslashbox{$m_H$}{$\sqrt{s}$} & 7 TeV & 8 TeV & 14 TeV & 100 TeV
   \\ \hline
   125 GeV & 13.8 & 17.9 & 39.6 & 583 \\ \hline
   300 GeV & 2.34 & 3.27 & 9.37 & 239 \\ \hline
  \end{tabular}
  \label{tbl:pp->h:NNLO}
 \end{subtable}
 \\
 \begin{subtable}{\textwidth}
  \caption{$\sigma^\text{NNLO}(pp \to h)$, pb, from Tables 1, 3
   of~\cite{handbook}.}
  \begin{tabular}{|c|c|c|c|c|}
   \hline
   \backslashbox{$m_H$}{$\sqrt{s}$} & 7 TeV & 8 TeV & 14 TeV & 100 TeV
   \\ \hline
   125 GeV & 15.31 & N/A & 49.97 & N/A \\ \hline
   300 GeV & 2.42  & N/A & 11.07 & N/A \\ \hline
  \end{tabular}
  \label{tbl:pp->h:NNLO-handbook}
 \end{subtable}
 \label{tbl:lhc-data}
\end{table}

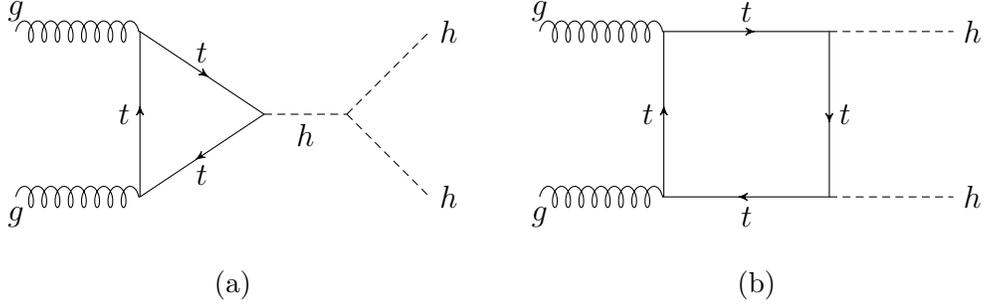
\begin{figure}[t]
 \centering
 \begin{subfigure}[b]{0.4\textwidth}
   \begin{tikzpicture}[scale=1.1]
     \coordinate (A) at (0,0);
     \coordinate (B) at (1.5,0);
     \coordinate (C) at (0,2);
     \coordinate (D) at (1.5,2);
     \coordinate (E) at (3,1);
     \coordinate (F) at (4,1);
     \coordinate (G) at (5,0);
     \coordinate (H) at (5,2);
     \draw[gluon] (A) node[below] {\normalsize $g$} -- (B);
     \draw[gluon] (C) node[above] {\normalsize $g$} -- (D);
     \draw[electron] (B) -- (D) node[midway,left] {\normalsize $t$};
     \draw[electron] (D) -- (E) node[midway,above] {\normalsize $t$};
     \draw[electron] (E) -- (B) node[midway,below] {\normalsize $t$};
     \draw[scalar] (E) -- (F) node[midway,below] {\normalsize $h$};
     \draw[scalar] (F) -- (G) node[right] {\normalsize $h$};
     \draw[scalar] (F) -- (H) node[right] {\normalsize $h$};
   \end{tikzpicture}
   \caption{}
   \label{fig:triangle}
 \end{subfigure}
 ~
 \begin{subfigure}[b]{0.4\textwidth}
   \begin{tikzpicture}[scale=1.1]
     \coordinate (A) at (0,0);
     \coordinate (B) at (1.5,0);
     \coordinate (C) at (0,2);
     \coordinate (D) at (1.5,2);
     \coordinate (E) at (3.5,0);
     \coordinate (F) at (3.5,2);
     \coordinate (G) at (5,0);
     \coordinate (H) at (5,2);
     \draw[gluon] (A) node[below] {\normalsize $g$} -- (B);
     \draw[gluon] (C) node[above] {\normalsize $g$} -- (D);
     \draw[electron] (B) -- (D) node[midway,left] {\normalsize $t$};
     \draw[electron] (D) -- (F) node[midway,above] {\normalsize $t$};
     \draw[electron] (E) -- (B) node[midway,below] {\normalsize $t$};
     \draw[electron] (F) -- (E) node[midway,right] {\normalsize $t$};
     \draw[scalar] (E) -- (G) node[right] {\normalsize $h$};
     \draw[scalar] (F) -- (H) node[right] {\normalsize $h$};
   \end{tikzpicture}
   \caption{}
   \label{fig:box}
 \end{subfigure}
\caption{Leading-order diagrams for the double higgs production at LHC.}
\label{fig:SMdiagrams}
\end{figure}

Let us turn now to double $h$ production at $pp$ collision in the SM. At the
leading order it is described by the two diagrams shown in
Fig.~\ref{fig:SMdiagrams}.  According to equations (4) and (11) and Table~1
from~\cite{contino}, the cross section of the double production of the 125~GeV
$h$ at 14~TeV LHC in the leading order equals:
\begin{equation}
 \sigma^\text{LO}(pp \to hh)
 = 144.6 \cdot \left( 0.169^2 + 0.457^2 - 1.79 \cdot 0.457 \cdot 0.169 \right)\text{ fb}
 = 14\text{ fb},
 \label{pp-hh-xsection-value}
\end{equation}
where the first term in parentheses originates from the square of the triangle
diagram, the second---from the square of the box diagram, while the last one is
their interference, which diminishes the cross section.

In order to understand result~\eqref{pp-hh-xsection-value} let us proceed in the
following way. In the limit $\hat s \ll 4 m_t^2$ the triangle $gg \to h$ and box
$gg \to hh$ amplitudes can be directly extracted from
lagrangian~\eqref{gg-h-amplitude}, 
expanding it over $h / v_\Phi$:
\begin{equation}
 \Delta \lagrangian
 = \frac{\alpha_s}{12 \pi}
   \ln \left(1 + \frac{h}{v_\Phi}\right)
   G_{\mu \nu}^2
 = \frac{\alpha_s}{12 \pi}
   \left( \frac{h}{v_\Phi} - \frac{1}{2} \frac{h^2}{v_\Phi^2} \right)
   G_{\mu \nu}^2,
\end{equation}
where the first term corresponds to the diagram shown in
Fig.~\ref{fig:triangle}, while the second term describes the diagram shown in
Fig.~\ref{fig:box}. Triple higgs coupling is given by the following term in the
SM lagrangian:
\begin{equation}
 \Delta \lagrangian = \frac{m_h^2}{2 v_\Phi} h^3,
\end{equation}
which leads to $\lambda_{hhh} = 3 m_h^3 / v_\Phi$. Hence, for the sum of the
triangle and the box diagrams at $\hat s \ll 4 m_t^2$ we get
\begin{equation}
 M
 = \frac{\alpha_s}{6 \pi v_\Phi}
   \left[
    \frac{1}{\hat s - m_h^2} \cdot 3 \frac{m_h^2}{v_\Phi} - \frac{1}{v_\Phi}
   \right]
   G_{\mu \nu}^1 G_{\mu \nu}^2,
\end{equation}
which equals zero at threshold when $\hat s = (2 m_h)^2$~\cite{spira, voloshin}.  For
the cross section we get
\begin{equation}
 \left. \hat \sigma_{gg \to hh} \right\rvert_{\hat s \ll 4 m_t^2}
 = \frac{\alpha_s^2 G_F^2 \hat s}{576 (2 \pi)^3}
   \left[ 1 - \frac{3 m_h^2}{\hat s - m_h^2} \right]^2
   \sqrt{1 - \frac{(2 m_h)^2}{\hat s}}
 \label{gg-hh-xsection-small-s}
\end{equation}
(see Eq.~13 from~\cite{spira}).

In the high-energy limit $\hat s \gg 4 m_t^2$ box diagram dominates and the
cross section behaves as:
\begin{equation}
 \left. \hat \sigma_{gg \to hh} \right\rvert_{\hat s \gg 4 m_t^2}
 = A^2 \frac{\alpha_s^2}{16 \pi^3 \hat s} \left( \frac{m_t}{v_\Phi} \right)^4
   \sqrt{1 - \frac{(2 m_h)^2}{\hat s}}.
 \label{gg-hh-xsection-big-s}
\end{equation}
Normalization constant $A$ is determined by the condition that at $\hat s = 4
m_t^2$ expressions~\eqref{gg-hh-xsection-small-s}
and~\eqref{gg-hh-xsection-big-s} are equal:
\begin{equation}
 A = \frac{1}{6} \left[ 1 - \frac{3 m_h^2}{4 m_t^2 - m_h^2} \right].
 \label{gg-hh-xsection-big-s-factor}
\end{equation}

\begin{figure}[h]
 \centering
 \input{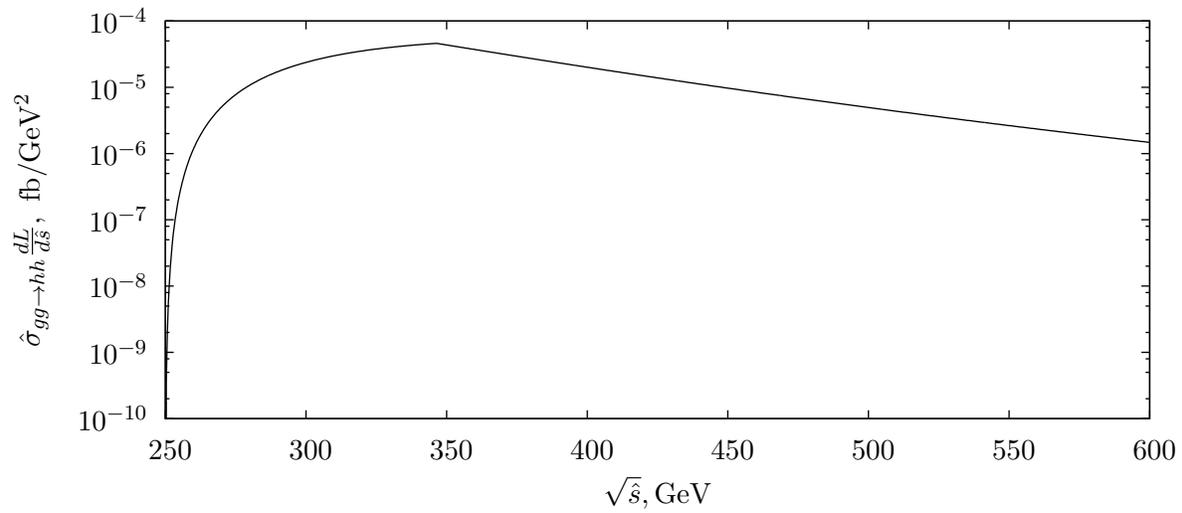}
 \caption{
   Differential cross section for the $pp \to hh$ reaction at $\sqrt{s} = 14$~TeV.
 }
 \label{fig:pp-hh-xsection}
\end{figure}

Finally, for the cross section of double $h$ production in the SM we obtain the
following approximate expression:
\begin{align}
 \sigma_{pp \to hh}
 &= \Int_{(2 m_h)^2}^s d \hat s \ 
     \hat \sigma_{gg \to hh}(\hat s) \frac{dL}{d \hat s},
 \label{pp-hh-xsection}
 \\
 \frac{dL}{d \hat s}
 &= \frac{1}{s}
    \Int_{\ln \sqrt{\tau}}^{-\ln \sqrt{\tau}}
     g(\sqrt{\tau} \e^y) g(\sqrt{\tau} \e^{-y}) dy,
\end{align}
where
Equations~\eqref{gg-hh-xsection-small-s}--\eqref{gg-hh-xsection-big-s-factor}
should be substituted in~\eqref{pp-hh-xsection} and $\tau \equiv \hat s / s$,
$\hat s$ being the $hh$ invariant mass. The differential cross section is shown
in Fig.~\ref{fig:pp-hh-xsection}, while for the total cross section for $hh$
production in the SM we get $\sigma(pp \to hh) = 4$~fb at $\sqrt{s} = 14$~TeV,
3.5 times smaller than the explicit leading order
result~\eqref{pp-hh-xsection-value}.

\clearpage
\section{Colored figures}

\vspace{-5mm}
\begin{minipage}{\textwidth}
 \centering
 \begin{minipage}{\textwidth}
  \centering
  \def\svgwidth{0.56\textwidth}
\begingroup%
  \makeatletter%
  \providecommand\color[2][]{%
    \errmessage{(Inkscape) Color is used for the text in Inkscape, but the package 'color.sty' is not loaded}%
    \renewcommand\color[2][]{}%
  }%
  \providecommand\transparent[1]{%
    \errmessage{(Inkscape) Transparency is used (non-zero) for the text in Inkscape, but the package 'transparent.sty' is not loaded}%
    \renewcommand\transparent[1]{}%
  }%
  \providecommand\rotatebox[2]{#2}%
  \ifx\svgwidth\undefined%
    \setlength{\unitlength}{567bp}%
    \ifx\svgscale\undefined%
      \relax%
    \else%
      \setlength{\unitlength}{\unitlength * \real{\svgscale}}%
    \fi%
  \else%
    \setlength{\unitlength}{\svgwidth}%
  \fi%
  \global\let\svgwidth\undefined%
  \global\let\svgscale\undefined%
  \makeatother%
  \begin{picture}(1,1)%
    \put(0,0){\includegraphics[width=\unitlength]{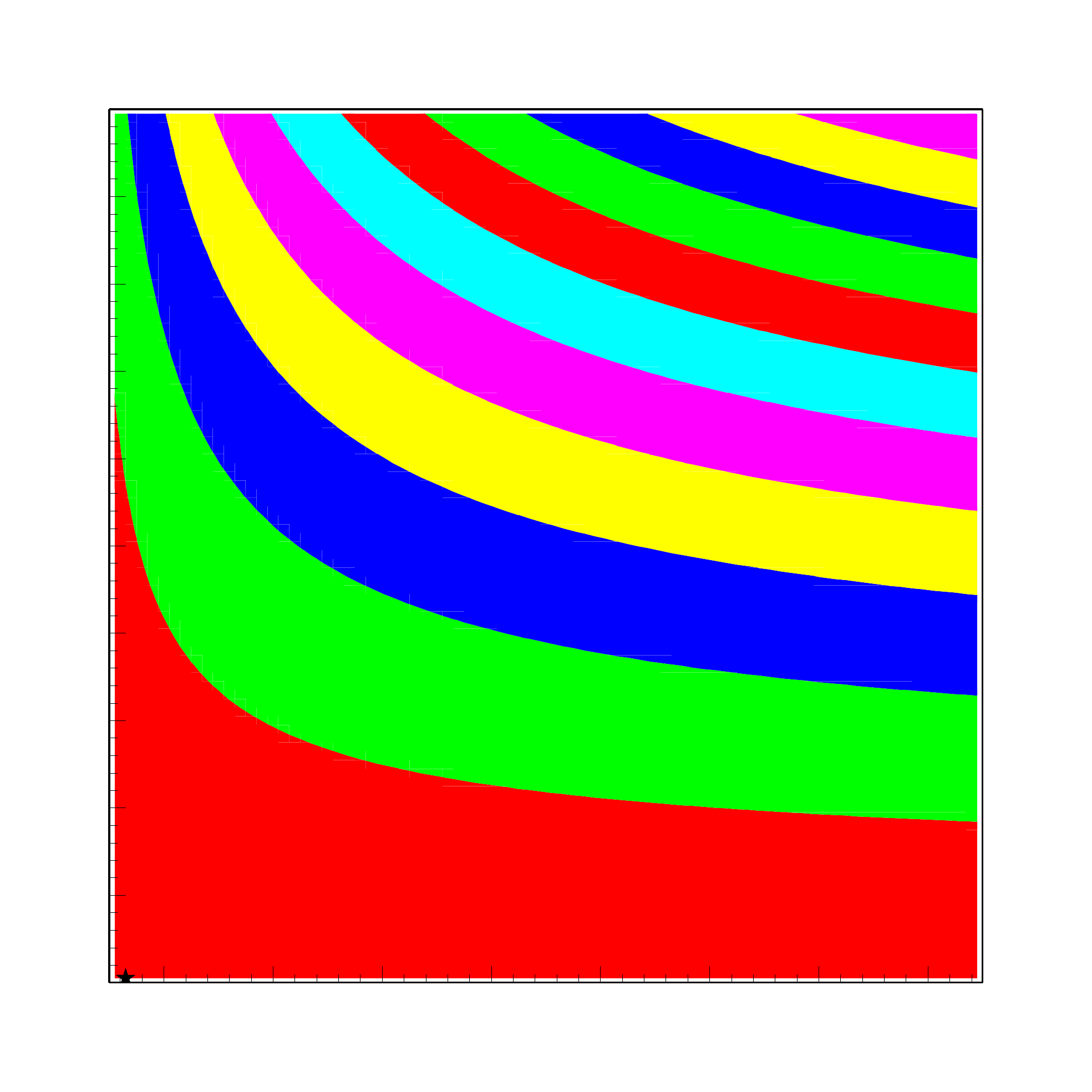}}%
    \put(0.1154661,0.05031779){\color[rgb]{0,0,0}\makebox(0,0)[lb]{\smash{200}}}%
    \put(0.21451271,0.05031779){\color[rgb]{0,0,0}\makebox(0,0)[lb]{\smash{300}}}%
    \put(0.31513902,0.05031779){\color[rgb]{0,0,0}\makebox(0,0)[lb]{\smash{400}}}%
    \put(0.41631356,0.05031779){\color[rgb]{0,0,0}\makebox(0,0)[lb]{\smash{500}}}%
    \put(0.51483052,0.05031779){\color[rgb]{0,0,0}\makebox(0,0)[lb]{\smash{600}}}%
    \put(0.61652542,0.05031779){\color[rgb]{0,0,0}\makebox(0,0)[lb]{\smash{700}}}%
    \put(0.8151483,0.05031779){\color[rgb]{0,0,0}\makebox(0,0)[lb]{\smash{900}}}%
    \put(0.71716099,0.05031779){\color[rgb]{0,0,0}\makebox(0,0)[lb]{\smash{800}}}%
    \put(0.4210587,0.00093917){\color[rgb]{0,0,0}\makebox(0,0)[lb]{\smash{$m_H$, GeV}}}%
    \put(0.02997835,0.16540996){\color[rgb]{0,0,0}\makebox(0,0)[lb]{\smash{0.1}}}%
    \put(0.02821492,0.24617516){\color[rgb]{0,0,0}\makebox(0,0)[lb]{\smash{0.2}}}%
    \put(0.02892029,0.32447156){\color[rgb]{0,0,0}\makebox(0,0)[lb]{\smash{0.3}}}%
    \put(0.02715686,0.40558946){\color[rgb]{0,0,0}\makebox(0,0)[lb]{\smash{0.4}}}%
    \put(0.02821492,0.48600196){\color[rgb]{0,0,0}\makebox(0,0)[lb]{\smash{0.5}}}%
    \put(0.02715686,0.56500373){\color[rgb]{0,0,0}\makebox(0,0)[lb]{\smash{0.6}}}%
    \put(0.02645149,0.64576896){\color[rgb]{0,0,0}\makebox(0,0)[lb]{\smash{0.7}}}%
    \put(0.02821492,0.72512338){\color[rgb]{0,0,0}\makebox(0,0)[lb]{\smash{0.8}}}%
    \put(0.02856761,0.80518321){\color[rgb]{0,0,0}\makebox(0,0)[lb]{\smash{0.9}}}%
    \put(0.02821492,0.88489033){\color[rgb]{0,0,0}\makebox(0,0)[lb]{\smash{1.0}}}%
    \put(-0.00079558,0.4440323){\color[rgb]{0,0,0}\rotatebox{90}{\makebox(0,0)[lb]{\smash{$\sin \alpha$}}}}%
    \put(0.90752819,0.23066569){\color[rgb]{0,0,0}\makebox(0,0)[lb]{\smash{$\Delta \chi^2 = 1\ (39\%)$}}}%
    \put(0.90752819,0.34583879){\color[rgb]{0,0,0}\makebox(0,0)[lb]{\smash{$\Delta \chi^2 = 4\ (86\%)$}}}%
    \put(0.90752819,0.43634028){\color[rgb]{0,0,0}\makebox(0,0)[lb]{\smash{$\Delta \chi^2 = 9\ (98.9\%)$}}}%
  \end{picture}%
\endgroup%

  \def\thefigure{\ref{fig:ew-bounds}}
  \captionof{figure}{Bounds from electroweak precision observables.}
 \end{minipage}
 \\
 \begin{minipage}{\textwidth}
  \centering
  \def\svgwidth{0.56\textwidth}
\begingroup%
  \makeatletter%
  \providecommand\color[2][]{%
    \errmessage{(Inkscape) Color is used for the text in Inkscape, but the package 'color.sty' is not loaded}%
    \renewcommand\color[2][]{}%
  }%
  \providecommand\transparent[1]{%
    \errmessage{(Inkscape) Transparency is used (non-zero) for the text in Inkscape, but the package 'transparent.sty' is not loaded}%
    \renewcommand\transparent[1]{}%
  }%
  \providecommand\rotatebox[2]{#2}%
  \ifx\svgwidth\undefined%
    \setlength{\unitlength}{567bp}%
    \ifx\svgscale\undefined%
      \relax%
    \else%
      \setlength{\unitlength}{\unitlength * \real{\svgscale}}%
    \fi%
  \else%
    \setlength{\unitlength}{\svgwidth}%
  \fi%
  \global\let\svgwidth\undefined%
  \global\let\svgscale\undefined%
  \makeatother%
  \begin{picture}(1,1)%
    \put(0,0){\includegraphics[width=\unitlength]{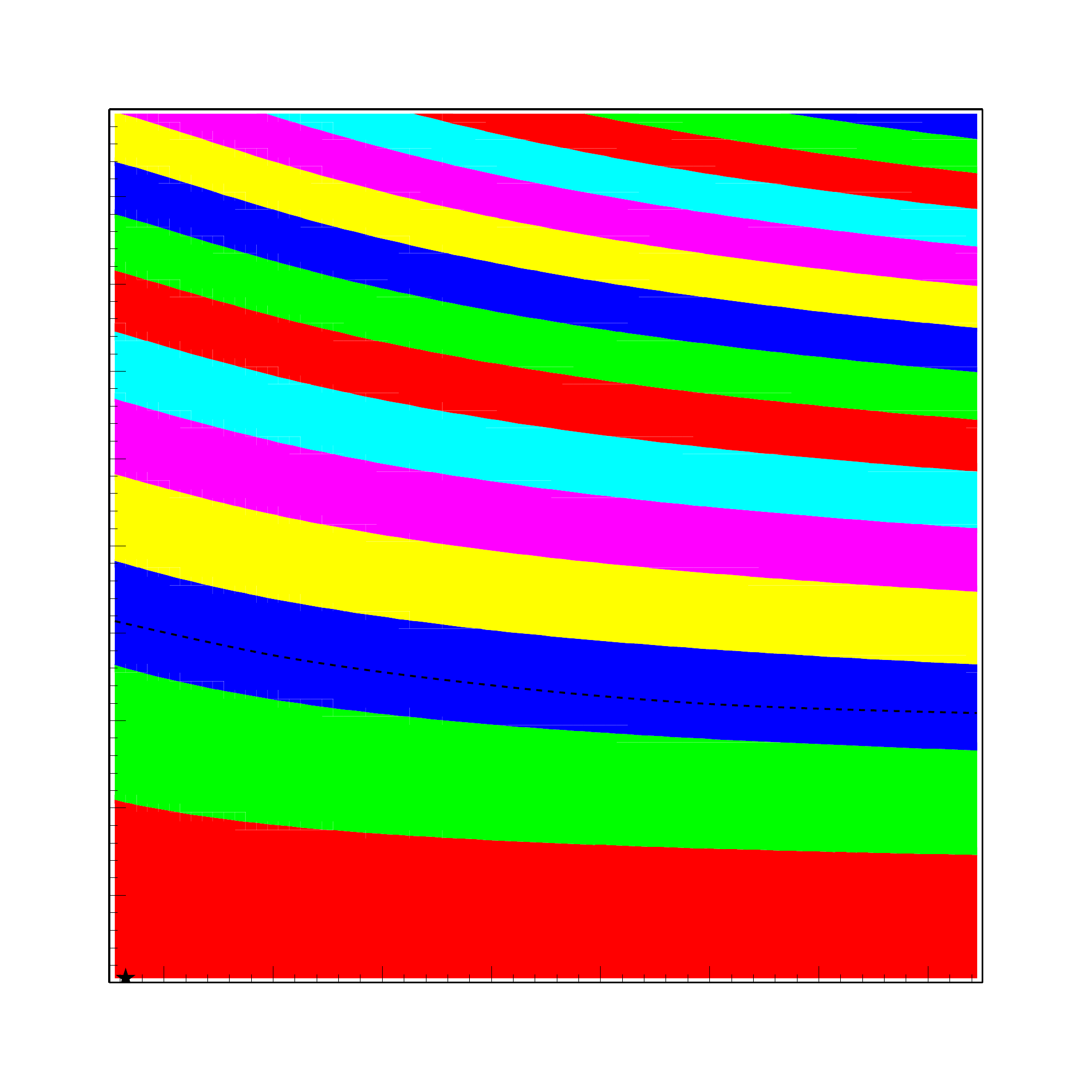}}%
    \put(0.81782168,0.04803915){\color[rgb]{0,0,0}\makebox(0,0)[lb]{\smash{900}}}%
    \put(0.11813952,0.04803915){\color[rgb]{0,0,0}\makebox(0,0)[lb]{\smash{200}}}%
    \put(0.21718612,0.04803915){\color[rgb]{0,0,0}\makebox(0,0)[lb]{\smash{300}}}%
    \put(0.31781243,0.04803915){\color[rgb]{0,0,0}\makebox(0,0)[lb]{\smash{400}}}%
    \put(0.41898694,0.04803915){\color[rgb]{0,0,0}\makebox(0,0)[lb]{\smash{500}}}%
    \put(0.5175039,0.04803915){\color[rgb]{0,0,0}\makebox(0,0)[lb]{\smash{600}}}%
    \put(0.61919881,0.04803915){\color[rgb]{0,0,0}\makebox(0,0)[lb]{\smash{700}}}%
    \put(0.71983442,0.04803915){\color[rgb]{0,0,0}\makebox(0,0)[lb]{\smash{800}}}%
    \put(0.4237321,-0.00133943){\color[rgb]{0,0,0}\makebox(0,0)[lb]{\smash{$m_H$, GeV}}}%
    \put(0.03265176,0.16313135){\color[rgb]{0,0,0}\makebox(0,0)[lb]{\smash{0.1}}}%
    \put(0.03088831,0.24389657){\color[rgb]{0,0,0}\makebox(0,0)[lb]{\smash{0.2}}}%
    \put(0.03159371,0.32219296){\color[rgb]{0,0,0}\makebox(0,0)[lb]{\smash{0.3}}}%
    \put(0.02983026,0.40331089){\color[rgb]{0,0,0}\makebox(0,0)[lb]{\smash{0.4}}}%
    \put(0.03088831,0.48372337){\color[rgb]{0,0,0}\makebox(0,0)[lb]{\smash{0.5}}}%
    \put(0.02983026,0.56272514){\color[rgb]{0,0,0}\makebox(0,0)[lb]{\smash{0.6}}}%
    \put(0.02912491,0.64349037){\color[rgb]{0,0,0}\makebox(0,0)[lb]{\smash{0.7}}}%
    \put(0.03088831,0.72284479){\color[rgb]{0,0,0}\makebox(0,0)[lb]{\smash{0.8}}}%
    \put(0.03124101,0.80290462){\color[rgb]{0,0,0}\makebox(0,0)[lb]{\smash{0.9}}}%
    \put(0.03088831,0.88261169){\color[rgb]{0,0,0}\makebox(0,0)[lb]{\smash{1.0}}}%
    \put(0.00187783,0.44175368){\color[rgb]{0,0,0}\rotatebox{90}{\makebox(0,0)[lb]{\smash{$\sin \alpha$}}}}%
    \put(0.90441346,0.20099587){\color[rgb]{0,0,0}\makebox(0,0)[lb]{\smash{$\Delta \chi^2 = 1\ (39\%)$}}}%
    \put(0.90441346,0.29956593){\color[rgb]{0,0,0}\makebox(0,0)[lb]{\smash{$\Delta \chi^2 = 4\ (86\%)$}}}%
    \put(0.90441346,0.38121245){\color[rgb]{0,0,0}\makebox(0,0)[lb]{\smash{$\Delta \chi^2 = 9\ (98.9\%)$}}}%
  \end{picture}%
\endgroup%

  \def\thefigure{\ref{fig:ew+mu-bounds}}
  \captionof{figure}{
   Bounds from both electroweak precision observables and signal strength
   measurements~\eqref{mu-atlas},~\eqref{mu-cms}. The dashed line corresponds to
   $\Delta \chi^2 = 5.99$; the probability that numerical values of $(m_H, \sin
   \alpha)$ are below it equals 95\% (compare with Ref.~\cite{lebedev}, eq.
   (23)).
  }
 \end{minipage}
 \def\thefigure{\ref{fig:ew-and-ew+mu-bounds}}
 \captionof{figure}{Bounds on the singlet model parameters.}
\end{minipage}

\begin{minipage}{\textwidth}
 \centering
 \begin{minipage}{\textwidth}
  \centering
  \def\svgwidth{0.56\textwidth}
\begingroup%
  \makeatletter%
  \providecommand\color[2][]{%
    \errmessage{(Inkscape) Color is used for the text in Inkscape, but the package 'color.sty' is not loaded}%
    \renewcommand\color[2][]{}%
  }%
  \providecommand\transparent[1]{%
    \errmessage{(Inkscape) Transparency is used (non-zero) for the text in Inkscape, but the package 'transparent.sty' is not loaded}%
    \renewcommand\transparent[1]{}%
  }%
  \providecommand\rotatebox[2]{#2}%
  \ifx\svgwidth\undefined%
    \setlength{\unitlength}{567bp}%
    \ifx\svgscale\undefined%
      \relax%
    \else%
      \setlength{\unitlength}{\unitlength * \real{\svgscale}}%
    \fi%
  \else%
    \setlength{\unitlength}{\svgwidth}%
  \fi%
  \global\let\svgwidth\undefined%
  \global\let\svgscale\undefined%
  \makeatother%
  \begin{picture}(1,1)%
    \put(0,0){\includegraphics[width=\unitlength]{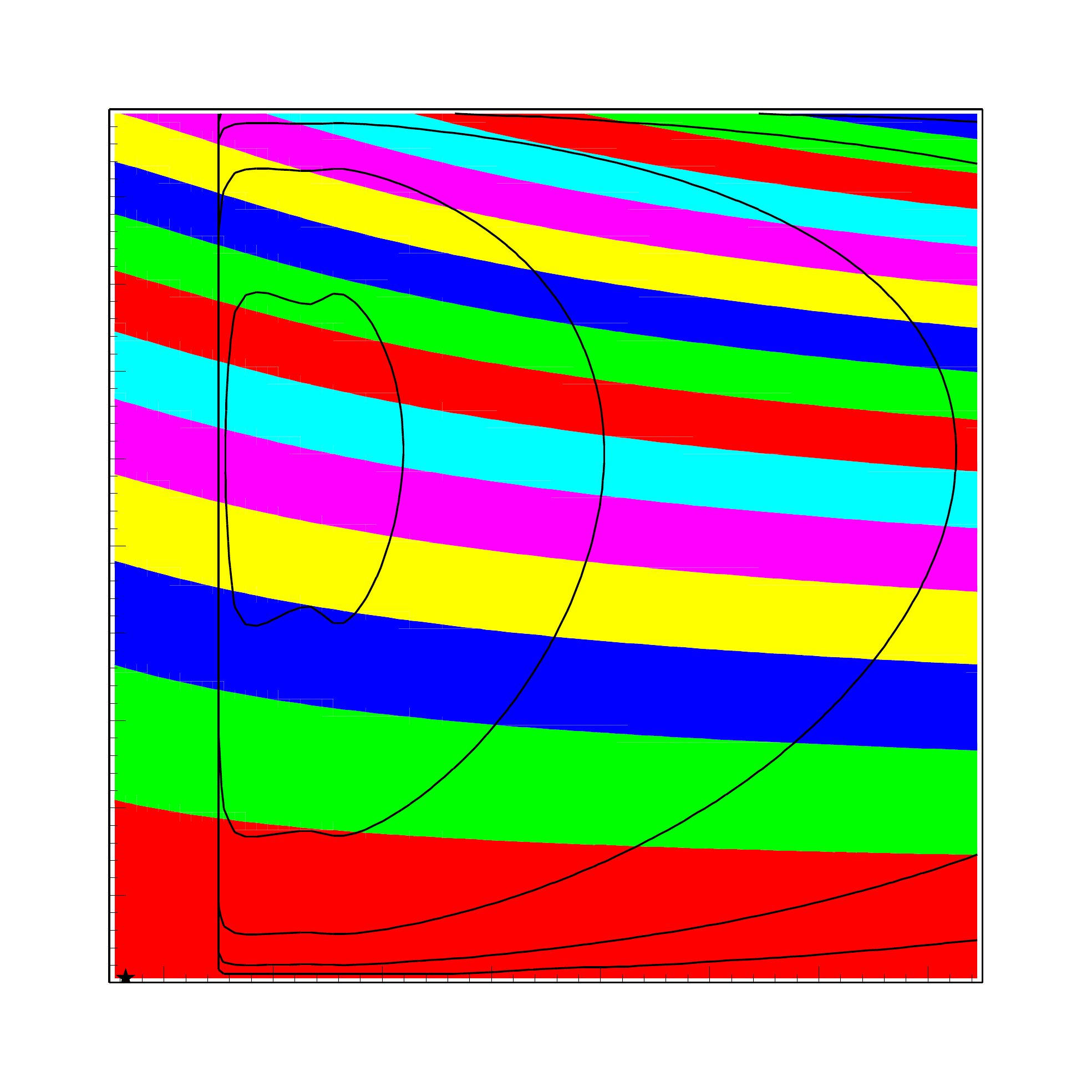}}%
    \put(0.81440743,0.05264591){\color[rgb]{0,0,0}\makebox(0,0)[lb]{\smash{900}}}%
    \put(0.11472526,0.05264591){\color[rgb]{0,0,0}\makebox(0,0)[lb]{\smash{200}}}%
    \put(0.21377191,0.05264591){\color[rgb]{0,0,0}\makebox(0,0)[lb]{\smash{300}}}%
    \put(0.3143982,0.05264591){\color[rgb]{0,0,0}\makebox(0,0)[lb]{\smash{400}}}%
    \put(0.41557272,0.05264591){\color[rgb]{0,0,0}\makebox(0,0)[lb]{\smash{500}}}%
    \put(0.5140897,0.05264591){\color[rgb]{0,0,0}\makebox(0,0)[lb]{\smash{600}}}%
    \put(0.61578461,0.05264591){\color[rgb]{0,0,0}\makebox(0,0)[lb]{\smash{700}}}%
    \put(0.71642017,0.05264591){\color[rgb]{0,0,0}\makebox(0,0)[lb]{\smash{800}}}%
    \put(0.42031788,0.00326733){\color[rgb]{0,0,0}\makebox(0,0)[lb]{\smash{$m_H$, GeV}}}%
    \put(0.02923754,0.16773812){\color[rgb]{0,0,0}\makebox(0,0)[lb]{\smash{0.1}}}%
    \put(0.02747408,0.24850335){\color[rgb]{0,0,0}\makebox(0,0)[lb]{\smash{0.2}}}%
    \put(0.02817948,0.32679972){\color[rgb]{0,0,0}\makebox(0,0)[lb]{\smash{0.3}}}%
    \put(0.02641603,0.40791765){\color[rgb]{0,0,0}\makebox(0,0)[lb]{\smash{0.4}}}%
    \put(0.02747408,0.48833012){\color[rgb]{0,0,0}\makebox(0,0)[lb]{\smash{0.5}}}%
    \put(0.02641603,0.5673319){\color[rgb]{0,0,0}\makebox(0,0)[lb]{\smash{0.6}}}%
    \put(0.02571068,0.64809712){\color[rgb]{0,0,0}\makebox(0,0)[lb]{\smash{0.7}}}%
    \put(0.02747408,0.72745155){\color[rgb]{0,0,0}\makebox(0,0)[lb]{\smash{0.8}}}%
    \put(0.02782678,0.80751137){\color[rgb]{0,0,0}\makebox(0,0)[lb]{\smash{0.9}}}%
    \put(0.02747408,0.88721844){\color[rgb]{0,0,0}\makebox(0,0)[lb]{\smash{1.0}}}%
    \put(-0.0015364,0.44636043){\color[rgb]{0,0,0}\rotatebox{90}{\makebox(0,0)[lb]{\smash{$\sin \alpha$}}}}%
    \put(0.78300069,0.13358001){\color[rgb]{0,0,0}\rotatebox{4.89973701}{\makebox(0,0)[lb]{\smash{0.1 fb}}}}%
    \put(0.65489728,0.15849787){\color[rgb]{0,0,0}\rotatebox{12.08850376}{\makebox(0,0)[lb]{\smash{1.0 fb}}}}%
    \put(0.36203504,0.15681857){\color[rgb]{0,0,0}\rotatebox{17.38222065}{\makebox(0,0)[lb]{\smash{0.01 pb}}}}%
    \put(0.21158613,0.24988127){\color[rgb]{0,0,0}\makebox(0,0)[lb]{\smash{0.1 pb}}}%
    \put(0.21670283,0.45638834){\color[rgb]{0,0,0}\makebox(0,0)[lb]{\smash{0.5 pb}}}%
    \put(0.90941737,0.19976765){\color[rgb]{0,0,0}\makebox(0,0)[lb]{\smash{$\Delta \chi^2 = 1\ (39\%)$}}}%
    \put(0.90941737,0.29833772){\color[rgb]{0,0,0}\makebox(0,0)[lb]{\smash{$\Delta \chi^2 = 4\ (86\%)$}}}%
    \put(0.90941737,0.37998424){\color[rgb]{0,0,0}\makebox(0,0)[lb]{\smash{$\Delta \chi^2 = 9\ (98.9\%)$}}}%
  \end{picture}%
\endgroup%

  \def\thefigure{\ref{fig:ew-bounds+pp->H->hh-xsection}}
  \captionof{figure}{
   Contour plot of $\sigma(pp \to H \to hh)$ for $\sqrt{s} = 14$~TeV.
   In this figure we neglect small effects of $H \to hh^*$.
  }
 \end{minipage}
 \\
 \begin{minipage}{\textwidth}
  \centering
  \def\svgwidth{0.56\textwidth}
\begingroup%
  \makeatletter%
  \providecommand\color[2][]{%
    \errmessage{(Inkscape) Color is used for the text in Inkscape, but the package 'color.sty' is not loaded}%
    \renewcommand\color[2][]{}%
  }%
  \providecommand\transparent[1]{%
    \errmessage{(Inkscape) Transparency is used (non-zero) for the text in Inkscape, but the package 'transparent.sty' is not loaded}%
    \renewcommand\transparent[1]{}%
  }%
  \providecommand\rotatebox[2]{#2}%
  \ifx\svgwidth\undefined%
    \setlength{\unitlength}{567bp}%
    \ifx\svgscale\undefined%
      \relax%
    \else%
      \setlength{\unitlength}{\unitlength * \real{\svgscale}}%
    \fi%
  \else%
    \setlength{\unitlength}{\svgwidth}%
  \fi%
  \global\let\svgwidth\undefined%
  \global\let\svgscale\undefined%
  \makeatother%
  \begin{picture}(1,1)%
    \put(0,0){\includegraphics[width=\unitlength]{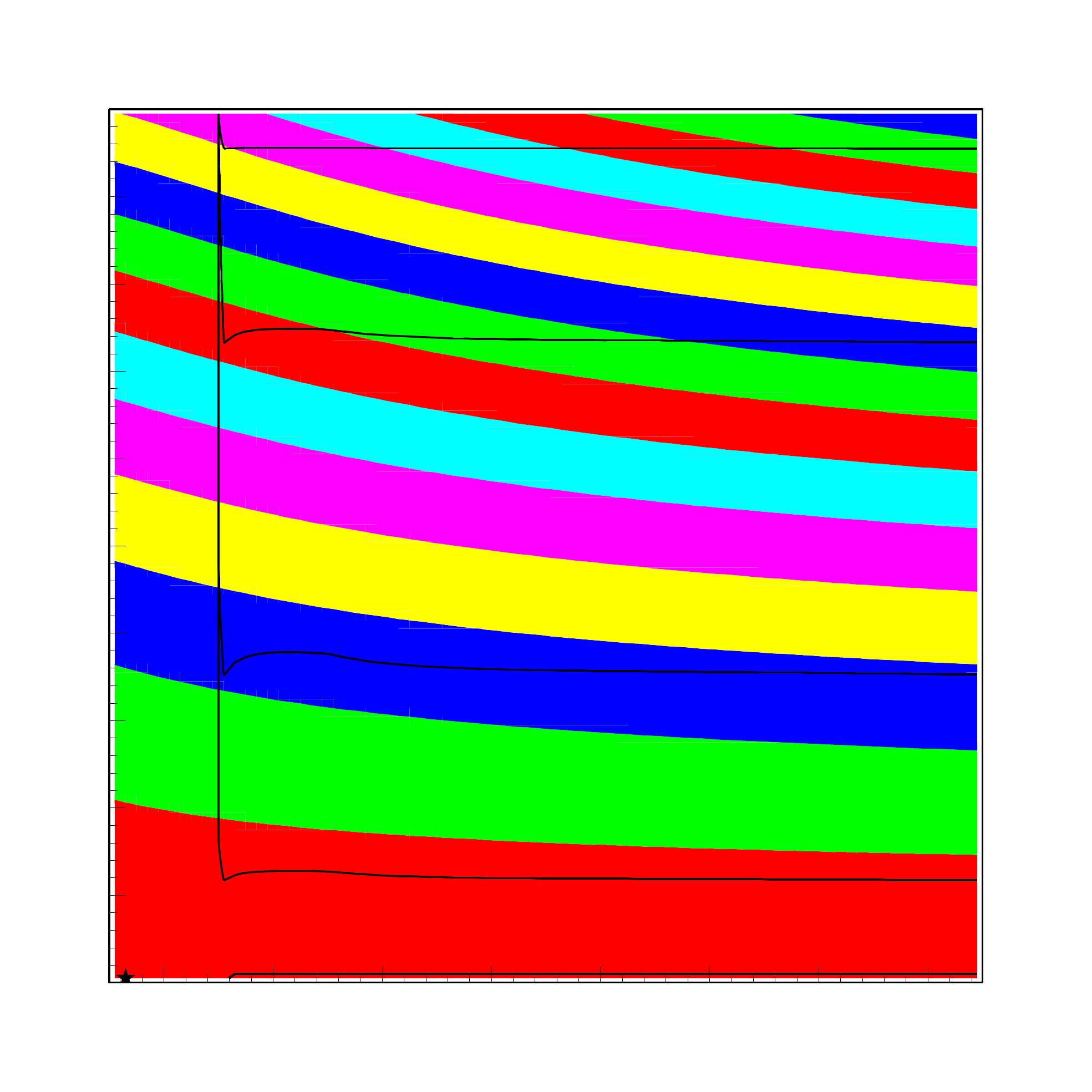}}%
    \put(0.8113258,0.05207723){\color[rgb]{0,0,0}\makebox(0,0)[lb]{\smash{900}}}%
    \put(0.11164368,0.05207723){\color[rgb]{0,0,0}\makebox(0,0)[lb]{\smash{200}}}%
    \put(0.2106903,0.05207723){\color[rgb]{0,0,0}\makebox(0,0)[lb]{\smash{300}}}%
    \put(0.31131662,0.05207723){\color[rgb]{0,0,0}\makebox(0,0)[lb]{\smash{400}}}%
    \put(0.41249111,0.05207723){\color[rgb]{0,0,0}\makebox(0,0)[lb]{\smash{500}}}%
    \put(0.51100807,0.05207723){\color[rgb]{0,0,0}\makebox(0,0)[lb]{\smash{600}}}%
    \put(0.61270298,0.05207723){\color[rgb]{0,0,0}\makebox(0,0)[lb]{\smash{700}}}%
    \put(0.71333865,0.05207723){\color[rgb]{0,0,0}\makebox(0,0)[lb]{\smash{800}}}%
    \put(0.41723633,0.00269864){\color[rgb]{0,0,0}\makebox(0,0)[lb]{\smash{$m_H$, GeV}}}%
    \put(0.02615593,0.16716944){\color[rgb]{0,0,0}\makebox(0,0)[lb]{\smash{0.1}}}%
    \put(0.02439247,0.24793463){\color[rgb]{0,0,0}\makebox(0,0)[lb]{\smash{0.2}}}%
    \put(0.02509788,0.326231){\color[rgb]{0,0,0}\makebox(0,0)[lb]{\smash{0.3}}}%
    \put(0.02333442,0.40734893){\color[rgb]{0,0,0}\makebox(0,0)[lb]{\smash{0.4}}}%
    \put(0.02439247,0.48776143){\color[rgb]{0,0,0}\makebox(0,0)[lb]{\smash{0.5}}}%
    \put(0.02333442,0.56676326){\color[rgb]{0,0,0}\makebox(0,0)[lb]{\smash{0.6}}}%
    \put(0.02262907,0.64752849){\color[rgb]{0,0,0}\makebox(0,0)[lb]{\smash{0.7}}}%
    \put(0.02439247,0.72688291){\color[rgb]{0,0,0}\makebox(0,0)[lb]{\smash{0.8}}}%
    \put(0.02474517,0.80694279){\color[rgb]{0,0,0}\makebox(0,0)[lb]{\smash{0.9}}}%
    \put(0.02439247,0.88664975){\color[rgb]{0,0,0}\makebox(0,0)[lb]{\smash{1.0}}}%
    \put(-0.00461804,0.44579171){\color[rgb]{0,0,0}\rotatebox{90}{\makebox(0,0)[lb]{\smash{$\sin \alpha$}}}}%
    \put(0.72215957,0.1604215){\color[rgb]{0,0,0}\makebox(0,0)[lb]{\smash{$R = 0.01$}}}%
    \put(0.72215957,0.35001582){\color[rgb]{0,0,0}\makebox(0,0)[lb]{\smash{$R = 0.1$}}}%
    \put(0.72215957,0.65336682){\color[rgb]{0,0,0}\makebox(0,0)[lb]{\smash{$R = 0.5$}}}%
    \put(0.72215957,0.82796999){\color[rgb]{0,0,0}\makebox(0,0)[lb]{\smash{$R = 0.9$}}}%
    \put(0.90183518,0.20239345){\color[rgb]{0,0,0}\makebox(0,0)[lb]{\smash{$\Delta \chi^2 = 1\ (39\%)$}}}%
    \put(0.90183518,0.30096347){\color[rgb]{0,0,0}\makebox(0,0)[lb]{\smash{$\Delta \chi^2 = 4\ (86\%)$}}}%
    \put(0.90183518,0.38260999){\color[rgb]{0,0,0}\makebox(0,0)[lb]{\smash{$\Delta \chi^2 = 9\ (98.9\%)$}}}%
  \end{picture}%
\endgroup%

  \def\thefigure{\ref{fig:zz-signal}}
  \captionof{figure}{
   Contour plot of
   $
    R \equiv \frac{ \sigma(pp \to H) \Br(H \to ZZ)}
                  {(\sigma(pp \to h) \Br(h \to ZZ))_\text{SM}}
   $. In the calculation of $R$ we assume $m_H > 2 m_h$.
  }
 \end{minipage}
\end{minipage}


\clearpage

\begin{thebibliography}{99}
 \bibitem{higgs-atlas}
  The ATLAS collaboration, Phys. Lett. {\bf B716} 1 (2012).
 \bibitem{higgs-cms}
  The CMS collaboration, Phys. Lett. {\bf B716} 30 (2012).
 \bibitem{dawson}
  Chien-Yi Chen, S. Dawson and I. M. Lewis, Phys. Rev. {\bf D91} (2015) 035015,
  arXiv:1410.5488.
 \bibitem{robens}
  T. Robens, T. Stefaniak, arXiv:1501.02234 (2015).
 \bibitem{martin-lozano}
  V. Martin-Lozano, J. M. Moreno, C. B. Park, arXiv:1501.03799 (2015).
 \bibitem{lebedev}
  A. Falkowski, C. Gross, O. Lebedev, arXiv:1502.01361 (2015).
 \bibitem{isotriplet}
  S. Godunov, M. Vysotsky, E. Zhemchugov, JETP Vol. 147 (3) (2015),
  arXiv:1408.0184.
 \bibitem{hill}
  A. Hill, J. J. van der Bij, Phys. Rev. {\bf D36}, 3463 (1987).
 \bibitem{atlas}
  The ATLAS collaboration, ATLAS-CONF-2014-009 (2014).
 \bibitem{cms}
  The CMS collaboration, CERN-PH-EP-2014-288, CMS-HIG-14-009, arXiv:1412.8662,
  (2014).
 \bibitem{leptop}
  V. A. Novikov, L. B. Okun, A. N. Rozanov, M. I. Vysotsky, CPPM-95-1,
  arXiv:hep-ph/9503308.
 \bibitem{pdf}
  L. A. Harland-Lang, A. D. Martin, P. Motylinki, R. S. Thorne, arXiv:1412.3989
  (2014).
 \bibitem{djouadi}
  J. Baglio, A. Djouadi, JHEP 1103 (2011) 055, arXiv:1012.0530.
 \bibitem{okun}
  L. B. Okun. {\it Leptons and Quarks.} World Scientific Publishing, Singapore,
  2014. ISBN 978-981-4603-00-3.
 \bibitem{handbook}
  S. Dittmaier, C. Mariotti, G. Passarino et al., CERN-2011-002,
  arXiv:1101.0593.
 \bibitem{florian}
  D. de Florian and J. Mazzitelli, PoS LL2014 (2014) 029, DESY 14-080 \slash{}
  LPN 14-073, arXiv:1405.4704.
 \bibitem{contino}
  R. Contino et al., JHEP 1208 (2012) 154, arXiv:1205.5444.
 \bibitem{spira}
  T. Plehn, M. Spira, and P. M. Zerwas. Nucl. Phys. B479, 46 (1996)
  [Erratum-ibid. B531, 655 (1996)], arXiv:hep-ph/9603205.
 \bibitem{voloshin}
  X. Li, M. B. Voloshin. Phys. Rev. {\bf D89} (2014) 1, 013012, arXiv:1311.5156.
 \bibitem{cms-bounds}
  The CMS collaboration, CMS PAS HIG-13-002 (2013).
 \bibitem{followups}
  S. Profumo, M. J. Ramsey-Musolf, G. Shaughnessy,
  JHEP 0708 (2007) 010, arXiv:0705.2425.
  \\
  V. Barger, P. Langacker, M. McCaskey, M. J. Ramsey-Musolf, G. Shaughnessy,
  Phys. Rev. {\bf D77} (2008) 035005, arXiv:0706.4311.
  \\
  V. Barger, P. Langacker, M. McCaskey, M. J. Ramsey-Musolf, G. Shaughnessy,
  Phys. Rev. {\bf D79} (2009) 015018, arXiv:0811.0393.
  \\
  M. Gonderinger, Y. Li, H. Patel, M. J. Ramsey-Musolf,
  JHEP 1001 (2010) 053, arXiv:0910.3167.
  \\
  M. Kadastik, K. Kannike, A. Racioppi, M. Raidal,
  JHEP 1205 (2012) 061, arXiv:1112.3647.
  \\
  M. Gonderinger, H. Lim, M. J. Ramsey-Musolf,
  Phys. Rev. {\bf D86} (2012) 043511, arXiv:1202.1316.
  \\
  C. Caillol, B. Clerbaux, J. M. Fr\`ere, S. Mollet,
  Eur. Phys. J. Plus 129 (2014) 93, arXiv:1304.0386.
  \\
  E. Gabrielli, M. Heikinheimo, K. Kannike et. al.,
  Phys. Rev. {\bf D89} (2014) 1, 015017, arXiv:1309.6632.
  \\
  L. Basso, O. Fischer, J. J. van der Bij,
  Phys. Lett. B730 (2014) 326-331, arXiv:1309.8096.
  \\
  J. M. No, M. J. Ramsey-Musolf,
  Phys. Rev. {\bf D89} (2014) 9, 095031, arXiv:1310.6035.
  \\
  J. de Blas, M. Chala, M. P\'erez-Victoria, J. Santiago,
  CERN-PH-TH-2014-264, arXiv:1412.8480 (2014).
  \\
  S. Profumo, M. J. Ramsey-Musolf, C. L. Wainwright, P. Winslow,
  Phys. Rev. {\bf D91} (2015) 3, 035018, arXiv:1407.5342.
  \\
  M. Gorbahn, J. M. No, V. Sanz,
  LTH 1039, arXiv:1502.07352 (2015).
  \\
  D. Curtin, P. Meade, Ch-T. Yu,
  JHEP 1411 (2014) 127, arXiv:1409.0005.
\end{thebibliography}
\end{document}